\documentstyle[12pt]{article}

\makeatletter
\renewcommand\thesubsection{\thesection.\@arabic\c@subsection}
\makeatother

\newcommand{\sect}[1]{\setcounter{equation}{0}\section{#1}}

\newcommand {\beq}{\begin{equation}}
\newcommand {\eeq}{\end{equation}}
\newcommand {\beqa}{\begin{eqnarray}}
\newcommand {\eeqa}{\end{eqnarray}}         %Equation version
\newcommand {\beqs}{\begin{eqnarray*}}
\newcommand {\eeqs}{\end{eqnarray*}}
\newcommand {\bds}{\begin{displaymath}}
\newcommand {\eds}{\end{displaymath}}
\newcommand {\n}{\nonumber\\}

\newcommand{\no}{\noindent}
\newcommand {\bebb}{\begin{thebibliography}}
\newcommand {\eebb}{\end{thebibliography}}      %Reference version
\newcommand {\bbit}{\bibitem}
\def\D{\Delta}

\def\gm{\gamma}

\def\L{\Lambda}

\def\p{\partial}

\def\lt{\left}
\def\rt{\right}
\def\dg{\dagger}

\def\journal#1&#2(#3){\unskip, \sl #1\ \bf #2 \rm(19#3) }
\def\andjournal#1&#2(#3){\sl #1~\bf #2 \rm (19#3) }

\def\npb#1#2#3{Nucl. Phys. {\bf B#1}, (#2) #3}

\def\plb#1#2#3{Phys. Lett. {\bf B#1}, (#2) #3}

\begin{document}

%\begin{titlepage}

\begin{flushright}
\end{flushright}

\vskip 1cm

\begin{center}
%\title
{\Large\bf Primary Fields and Screening Currents of $gl(2|2)$ Non-unitary 
Conformal Field Theory} 

\vspace{1cm}

%\author{
{\large Yao-Zhong Zhang $^a$, Xin Liu $^a$ and Wen-Li Yang $^{a,b}$}
\vskip.1in
$a.$ {\em Department of Mathematics, 
University of Queensland, Brisbane, Qld 4072, Australia}

$b.$ {\em Institute of Modern Physics, Northwest University, 
Xi'an 710069, China}

\end{center}

\date{}

%\maketitle

%\vspace{2cm}

\begin{abstract}
The non-semisimple $gl(2|2)_k$ current superalgebra in the standard
basis and the corresponding non-unitary conformal field theory are 
investigated. Infinite families of primary fields corresponding 
to all finite-dimensional
irreducible typical and atypical representations of $gl(2|2)$ and 
three (two even and one odd) screening currents of the first kind are 
constructed explicitly in terms of ten free fields.
\end{abstract}

\vskip.1in

{\it PACS}: 11.25Hf. % 11.30.Rd; 03.65Fd; 02.20.Hj.

{\it Keywords}: Superalgebra; Conformal field theory; Free
field realizations.

%\vspace{0.5cm}

%\end{titlepage}

\setcounter{section}{0}
\setcounter{equation}{0}
\sect{Introduction}

Current superalgebras with zero superdimension and their
corresponding conformal field theories (CFTs) have emerged
in a wide variety of areas ranging from theoretical high energy 
physics to condensed matter
physics. In particular, they have found important applications in
topological field theories \cite{Roz92,Isi94}, logarithmic CFTs (see
e.g. \cite{Flohr} and references therein), disordered systems and the
integer quantum Hall transition
\cite{Ber95}-\cite{Gur00}, and sigma models on supergroup
manifolds \cite{Ber99}, to mention just a few. One of the
interesting characterizations of such superalgebra CFTs is the vanishing of
Virasoro central charges and the existence of primary fields with negative
conformal dimensions which render the CFTs non-unitary. The non-unitarity makes
the CFTs non-trivial though their Virasoro central charges are zero.

It is well-known that unlike a bosonic algebra most superalgebras admit
different Weyl inequivalent choices of simple root systems, which
correspond to many inequivalent Dynkin diagrams (see e.g. \cite{Fra96}
and references therein). Given a basic superalgebra,
there exists a particular simple root system which contains the smallest
number of odd roots. Such a simple root system is called distinguished
or standard. Otherwise the simple root system is called non-standard.
Very often one needs to work in different bases (i.e. different simple
root systems) for different physical applications.

The $osp(2|2)_k$ current superalgebra in both the standard and non-standard
bases and the corresponding CFTs were investigated in 
\cite{Ito}-\cite{Zha03a} by means of the free field
approach. The Wakimoto realization of the $sl(2|2)_k$ current superalgebra in
the standard basis and that of the $gl(2|2)_k$ current superalgebra 
in the non-standard basis were constructed in \cite{Ras01}
and \cite{Din03b}, respectively. Three fermionic screen currents for the
latter case were also obtained in \cite{Din03b}.

In this paper we investigate the $gl(2|2)_k$ current superalgebra and
the corresponding non-unitary CFT in the standard basis. 
In sections 3 and 4, we construct explicitly  the Sugawara energy-momentum
tensor and infinite families of primary fields of the $gl(2|2)$ CFT
for arbitrary level $k$. This forms the first
main new result of this paper. We moreover construct, in section 5,
three (two even and one odd) screening current operators
of the first kind in the standard basis. To our knowledge, such explicit
construction of screening currents is also new. In section 6, we give a
brief discussion on the correlation functions and derive the
Knizhnik-Zamolodchikov equation satisfied by these correlators.

Let us mention three new and interesting features of our results which
do not appear in ordinary CFTs associated with bosonic current algebras.
i)  The operator product expansions (OPEs) of some affine supercurrents
depend on a free
parameter. ii) This free parameter also appears in the Sugawara 
energy-momentum tensor in the free field realization. iii) Primary
fields depend on two arbitrary parameters for typical representations and
one arbitrary parameter for atypical representations. It follows that
to a given representation of $gl(2|2)$ there corresponds an infinite
family of primary fields.

\sect{$gl(2|2)_k$ current superalgebra and free field realization}

The superalgebra $gl(2|2)$ is non-semisimple and can be written as 
$gl(2|2)=gl(2|2)^{\rm even}\oplus gl(2|2)^{\rm odd}$, where
\beqa
gl(2|2)^{\rm even}&=&gl(2)\oplus gl(2)\n
&=&\{H'\}\oplus\{\{E_{12},E_{21},H_1\}\oplus\{E_{34},E_{43},H_2\},H_{ex}\},\n
gl(2|2)^{\rm odd}&=&\{E_{13},E_{31},E_{23},E_{32},E_{24},E_{42},
   E_{14},E_{41}\}.
\eeqa
In the standard basis, $E_{12},E_{34},E_{23}$
($E_{21},E_{43},E_{32}$) are the simple raising (lowering) generators, 
$E_{13},E_{14},E_{24}$ ($E_{31},E_{41},E_{42}$) are the non-simple
raising (lowering) generators and $H_1,H_2,H',H_{ex}$ are the elements of
the Cartan subalgebra. We have  
\beqa
H_1&=&E_{11}-E_{22},~~~~H_2=E_{33}-E_{44},\n
H'&=&E_{11}+E_{22}+E_{33}+E_{44},\n
H_{ex}&=&E_{11}+E_{22}-E_{33}-E_{44}+\beta H'\label{cartan-gl22}
\eeqa
with $\beta$ being an arbitrary parameter. That $H_{ex}$ is not uniquely
determined is a consequence of the fact that $gl(2|2)$ is
non-semisimple. The generators obey the following (anti-)commutation
relations:
\beq
[E_{ij},E_{kl}]=\delta_{jk}E_{il}-(-1)^{([i]+[j])([k]+[l])}\delta_{il}
   E_{kj},
\eeq
where $[E_{ij},E_{kl}]\equiv
E_{ij}E_{kl}-(-1)^{([i]+[j])([k]+[l])}E_{kl}E_{ij}$ is a commutator or
an anticommutator, $[1]=[2]=0,~
[3]=[4]=1$, and $E_{ii},~i=1,2,3,4$, are related to $H_1,H_2,H',H_{ex}$
via (\ref{cartan-gl22}). 

The quadratic Casimir of $gl(2|2)$ is given by 
$C_1=\sum_{AB}\,(-1)^{[B]}E_{AB}E_{BA}$. Because $gl(2|2)$ is
non-semisimple, $\sum_{A}\,E_{AA}$ is a central element and thus there
is another quadratic Casimir $C_2=\sum_{AB}\,E_{AA}E_{BB}=\lt(H'\rt)^2$.
Note that $C_1$ and $C_2$ are two independent quadratic Casimirs; in
the defining representation, $C_1$ actually vanishes and $C_2$ is just
the identity matrix. These two Casimir elements are useful in the following
for the construction of the Sugawara energy-momentum tensor.

We shall use $E_{ij}(z)$ to denote the (affine)
supercurrents associated with the $gl(2|2)$-generators $E_{ij}$.
Then the $gl(2|2)$ current superalgebra at arbitrary level $k$ has the
form
\beq
E_{ij}(z)E_{kl}(w)=k\frac{{\rm str}(E_{ij}E_{kl})}{(z-w)^2}
  +\frac{1}{z-w}\lt(\delta_{jk}E_{il}(w)-(-1)^{([i]+[j])([k]+[l])}
  \delta_{il}E_{kj}(w)\rt),
\eeq
where 
\beqa
E_{11}(z)&=&\frac{1}{4}[(1-\beta)H'(z)+H_{ex}(z)+2H_1(z)],\n
E_{22}(z)&=&\frac{1}{4}[(1-\beta)H'(z)+H_{ex}(z)-2H_1(z)],\n
E_{33}(z)&=&\frac{1}{4}[(1+\beta)H'(z)-H_{ex}(z)+2H_2(z)],\n
E_{44}(z)&=&\frac{1}{4}[(1+\beta)H'(z)-H_{ex}(z)-2H_2(z)].
\eeqa

The Wakimoto free field realization of the $sl(2|2)_k$ current superalgebra
in the standard basis was given in \cite{Ras01}, while that of 
the non-semisimple $gl(2|2)_k$ current superalgebra
in the non-standard basis was obtained in \cite{Din03b}. Here we 
give a realization of the $gl(2|2)_k$ current superalgebra in
the standard basis by extending the result in \cite{Ras01} through
the addition of the current $H_{ex}(z)$. 
We introduce two bosonic $\beta$-$\gamma$ pairs, four
fermionic $b$-$c$ type systems and four free scalar fields
$\phi_1(z), \phi_2(z), \phi'(z)$ and $\phi_{ex}(z)$, which
have the following OPEs:
\beqa
&&\beta_{ij}(z)\gamma_{kl}(w)=-\gamma_{kl}(z)\beta_{ij}(w)=\frac{\delta_{ik}
   \delta_{jl}}{z-w},\n
&&\psi^\dg_{ij}(z)\psi_{kl}(w)=\psi_{kl}(z)\psi^\dg_{ij}(w)=\frac{\delta_{ik}
   \delta_{jl}}{z-w},\n
&&\phi_1(z)\phi_1(w)=2\ln(z-w),~~~\phi_2(z)\phi_2(w)=-2\ln(z-w),\n
&&\phi'(z)\phi_{ex}(w)=4\ln(z-w),~~~\phi_{ex}(z)\phi_{ex}(w)=8\beta\ln(z-w),
\eeqa
and all other OPEs between the free fields are trivial.
Then the free field realization in the standard basis at arbitrary level
$k$ reads
\beqa
E_{12}(z)&=&\beta_{12}(z)-\frac{1}{2}\psi_{23}(z)\psi_{13}^\dg(z)
   -\frac{1}{2}\lt[\psi_{24}(z)-\frac{1}{6}\gamma_{34}(z)\psi_{23}(z)
   \rt]\psi_{14}^\dg(z),\n
E_{34}(z)&=&\beta_{34}(z)+\frac{1}{2}\psi_{23}(z)\psi_{24}^\dg(z)
   +\frac{1}{2}\lt[\psi_{13}(z)+\frac{1}{6}\gamma_{12}(z)\psi_{23}(z)
   \rt]\psi_{14}^\dg(z),\n
E_{23}(z)&=&\psi_{23}^\dg(z)+\frac{1}{2}\gm_{12}(z)\psi_{13}^\dg(z)
   -\frac{1}{2}\gm_{34}(z)\lt[\psi_{24}^\dg(z)+\frac{1}{3}\gm_{12}(z)
   \psi_{14}^\dg(z)\rt],\n
H_1(z)&=&\sqrt{k}\p\phi_1(z)-2\gm_{12}(z)\beta_{12}(z)+\psi_{23}(z)
   \psi_{23}^\dg (z)-\psi_{13}(z)\psi_{13}^\dg(z)\n
& &   +\psi_{24}(z)\psi_{24}^\dg(z) -\psi_{14}(z)\psi_{14}^\dg(z),\n
H_2(z)&=&\sqrt{k}\p\phi_2(z)-2\gm_{34}(z)\beta_{34}(z)+\psi_{23}(z)
   \psi_{23}^\dg (z)+\psi_{13}(z)\psi_{13}^\dg(z)\n
& &  -\psi_{24}(z)\psi_{24}^\dg(z) -\psi_{14}(z)\psi_{14}^\dg(z),\n
H'(z)&=&\sqrt{k}\p\phi'(z),\n
H_{ex}(z)&=&\sqrt{k}\p\phi_{ex}(z)-\frac{2}{\sqrt{k}}\p\phi'(z)\n
& &   -2\lt[\psi_{23}(z)\psi_{23}^\dg(z)+\psi_{13}(z)\psi_{13}^\dg(z)
    +\psi_{24}(z)\psi_{24}^\dg(z)+\psi_{14}(z)\psi_{14}^\dg(z)\rt],\n
E_{21}(z)&=&\gm_{12}(z)[\sqrt{k}\p\phi_1(z)-\gm_{12}(z)\beta_{12}(z)]\n
& &  +\frac{1}{2}\gm_{12}(z)\lt[\psi_{23}(z)\psi_{23}^\dg(z)-
   \psi_{13}(z)\psi_{13}^\dg(z)+\psi_{24}(z)\psi_{24}^\dg(z)
   -\psi_{14}(z)\psi_{14}^\dg(z)\rt]\n
& &-\psi_{13}(z)\psi_{23}^\dg(z)-\psi_{14}(z)\psi_{24}^\dg(z)
   -\frac{1}{4}(\gm_{12}(z))^2\lt[\psi_{23}(z)\psi_{13}^\dg(z)
   +\psi_{24}(z)\psi_{14}^\dg(z)\rt]\n
& &+\frac{1}{12}\gm_{12}(z)\gm_{34}(z)\lt[\psi_{23}(z)\psi_{24}^\dg(z)
   -\psi_{13}(z)\psi_{14}^\dg(z)\rt]+(k-1)\p\gm_{12}(z),\n
E_{43}(z)&=&\gm_{34}(z)[\sqrt{k}\p\phi_2(z)-\gm_{34}(z)\beta_{34}(z)]\n
& &  +\frac{1}{2}\gm_{34}(z)\lt[\psi_{23}(z)\psi_{23}^\dg(z)
   +\psi_{13}(z)\psi_{13}^\dg(z)-\psi_{24}(z)\psi_{24}^\dg(z)
   -\psi_{14}(z)\psi_{14}^\dg(z)\rt]\n
& &+\psi_{24}(z)\psi_{23}^\dg(z)+\psi_{14}(z)\psi_{13}^\dg(z)
   +\frac{1}{4}(\gm_{34}(z))^2\lt[\psi_{23}(z)\psi_{24}^\dg(z)
   +\psi_{13}(z)\psi_{14}^\dg(z)\rt]\n
& &+\frac{1}{12}\gm_{12}(z)\gm_{34}(z)\lt[\psi_{24}(z)\psi_{14}^\dg(z)
   -\psi_{23}(z)\psi_{13}^\dg(z)\rt]-(k+1)\p\gm_{34}(z),\n
E_{32}(z)&=&\psi_{23}(z)\frac{1}{2}\sqrt{k}[\p\phi'(z)-\p\phi_1(z)
   +\p\phi_2(z)]\n
& &+\frac{1}{2}\psi_{23}(z)\lt[\gm_{12}(z)\beta_{12}(z)-\gm_{34}(z)
   \beta_{34}(z)+\psi_{13}(z)\psi_{13}^\dg(z)
   -\psi_{24}(z)\psi_{24}^\dg(z)\rt]\n 
& &+\psi_{13}(z)\beta_{12}(z)+\psi_{24}(z)\beta_{34}(z)\n
& &+\frac{1}{6}\psi_{23}(z)\lt[\gm_{12}(z)\psi_{24}(z)+\gm_{34}(z)
   \psi_{13}(z)\rt]\psi_{14}^\dg(z)+k\p\psi_{23}(z).
\eeqa

Some remarks are in order.
There appears an ``anomalous" term, $-\frac{2}{\sqrt{k}}\p\phi'(z)$,
in the current $H_{ex}(z)$.
Moreover, although the parameter $\beta$ does not occur in the
expression of the supercurrents, it appears in the OPE of $H_{ex}(z)$
with itself:
\beq
H_{ex}(z)H_{ex}(w)=k\frac{8\beta}{(z-w)^2}.
\eeq
So different choice of $\beta$ will give different OPEs of $H_{ex}(z)$
with itself. This is one of the characterizations of the non-semisimple
$gl(2|2)_k$ current superalgebra.

\sect{Energy-momentum tensor in the standard basis}

The CFT associated with the $gl(2|2)_k$ current superalgebra is obtained
by constructing the Sugawara energy-momentum tensor. Such a tensor
corresponding to $gl(2|2)$ in the non-standard basis has been obtained
in \cite{Din03b}. Here we construct the Sugawara tensor in the standard
basis. Keeping in mind
that the dual Coxeter number of $gl(2|2)$ is zero, the Sugawara tensor
corresponding to the quadratic Casimir $C_1$ is given by
\beqa
T_1(z)&=&\frac{1}{2k}:\sum_{i,j}(-1)^{[j]}E_{ij}(z)E_{ji}(z):\n
&=&\beta_{12}(z)\p\gm_{12}(z)+\beta_{34}(z)\p\gm_{34}(z)\n
& & -\psi_{13}^\dg(z)\p\psi_{13}(z)-\psi_{23}^\dg(z)\p\psi_{23}(z)
  -\psi_{14}^\dg(z)\p\psi_{14}(z)-\psi_{24}^\dg(z)\p\psi_{24}(z)\n
& &+\frac{1}{4}\lt[(\p\phi_1(z))^2-(\p\phi_2(z))^2
  +\p\phi'(z)\p\phi_{ex}(z)-\frac{2+\beta k}{k}(\p\phi'(z))^2\rt]\n
& &-\frac{1}{2\sqrt{k}}\lt[\p^2\phi_1(z) -\p^2\phi_2(z) -2\,\p^2\phi'(z)\rt].
\eeqa
With respect to this tensor, all supercurrents except $H_{ex}(z)$ are
primary fields. However, its OPE with $H_{ex}(z)$ reads
\beq
T_1(z)H_{ex}(w)=\frac{H_{ex}(w)}{(z-w)^2}+\frac{\p H_{ex}(w)}{z-w}
   -\frac{4/\sqrt{k}}{(z-w)^2}\p\phi'(w)
   -\frac{4/\sqrt{k}}{z-w}\p^2\phi'(w).\label{t1-Hex}
\eeq
This implies that $T_1(z)$ is not a correct energy-momentum tensor of
the $gl(2|2)$ CFT in the standard basis. This is another important
characterization of the non-semisimple $gl(2|2)_k$ current superalgebra.

Now the Sugawara tensor associated with the Casimir $C_2$ is
\beq
T_2(z)=\frac{1}{2k}:\sum_{i,j}E_{ii}(z)E_{jj}(z):=\frac{1}{2}(\p\phi'(z))^2.
\eeq
It has a non-trivial OPE with the current $H_{ex}(z)$:
\beq
T_2(z)H_{ex}(w)=\frac{4\sqrt{k}}{(z-w)^2}\p\phi'(w)
  +\frac{4\sqrt{k}}{z-w}\p^2\phi'(w).\label{t2-Hex}
\eeq
Comparing the OPEs (\ref{t1-Hex}) and (\ref{t2-Hex}), we find that 
if one defines
\beq
T(z)=T_1(z)+\frac{1}{k}T_2(z)
\eeq
then all currents become primary with respect to $T(z)$ and moreover
\beq
T(z)T(w)=\frac{c/2}{(z-w)^4}+\frac{2T(w)}{(z-w)^2}+\frac{\p T(w)}{z-w},
\eeq
where the central charge $c=0$. So $T(z)$ is the energy-momentum tensor
of the $gl(2|2)_k$ current superalgebra in the standard basis. In terms 
of free fields, $T(z)$ reads
\beqa
T(z)&=&\beta_{12}(z)\p\gm_{12}(z)+\beta_{34}(z)\p\gm_{34}(z)\n
& & -\psi_{13}^\dg(z)\p\psi_{13}(z)-\psi_{23}^\dg(z)\p\psi_{23}(z)
  -\psi_{14}^\dg(z)\p\psi_{14}(z)-\psi_{24}^\dg(z)\p\psi_{24}(z)\n
& &+\frac{1}{4}\lt[(\p\phi_1(z))^2-(\p\phi_2(z))^2
  +\p\phi'(z)\p\phi_{ex}(z)-\beta (\p\phi'(z))^2\rt]\n
& &-\frac{1}{2\sqrt{k}}\lt[\p^2\phi_1(z) -\p^2\phi_2(z) -2\,\p^2\phi'(z)\rt].
\eeqa
Note that the energy-momentum tensor depends on the free paramter
$\beta$.

\sect{Primary fields in the standard basis}\label{primary-fields}

Unlike ordinary bosonic algebras, there are two types of representations
for most superalgebras, i.e. the so-called typical and atypical
representations. The typical representations are irreducible and similar
to the usual representations appeared in ordinary bosonic algebras. The
atypical representations have no counterpart in the bosonic algebra
setting. They can be irreducible or not fully reducible (i.e.
indecomposable).

The main new result in this paper is the explicit construction of
primary fields associated with all finite-dimensional irreducible typical and
atypical representations of $gl(2|2)$.

Primary fields are fundamental objects in CFTs. A
primary field $\Psi$ has the following OPE with the energy-momentum tensor:
\beq
T(z)\Psi(w)=\frac{\Delta_\Psi}{(z-w)^2}\Psi(w)+\frac{\p\Psi(w)}{z-w},
   \label{primary-def}
\eeq
where $\Delta_\Psi$ is the conformal dimension of $\Psi(z)$. Moreover,
the OPEs of $\Psi(z)$ with affine currents do not contain poles higher
than first order. A special type of primary field is the highest weight
field (i.e. the vertex operator).

We now construct primary fields of the $gl(2|2)$ CFT in the standard
basis for general level $k$. It can be shown that the vertex operator given by
\beq
V_{J_1,J_2,q,p}(z)=\exp\lt\{\frac{1}{\sqrt{k}}[J_1\phi_1(z)-J_2\phi_2(z)]
   +\frac{q}{2\sqrt{k}}[\phi_{ex}(z)+\frac{2}{k}\phi'(z)]
   +\frac{p-2\beta q}{2\sqrt{k}}\phi'(z)\rt\},
\eeq
where $J_1,J_2$ are non-negative integer or half-interger and $q,p$ 
are arbitrary complex numbers, is the highest
weight field of the $gl(2|2)_k$ current superalgebra with highest weight
$(J_1,J_2,q,p)$. That is,
\beqa
&&H_1(z)V_{J_1,J_2,q,p}(w)=\frac{2J_1}{z-w}V_{J_1,J_2,q,p}(w),~~~~
H_2(z)V_{J_1,J_2,q,p}(w)=\frac{2J_2}{z-w}V_{J_1,J_2,q,p}(w),\n
&&H'(z)V_{J_1,J_2,q,p}(w)=\frac{2q}{z-w}V_{J_1,J_2,q,p}(w),~~~~
H_{ex}(z)V_{J_1,J_2,q,p}(w)=\frac{2p}{z-w}V_{J_1,J_2,q,p}(w),\n
&&E_{12}(z)V_{J_1,J_2,q,p}(w)=E_{34}(z)V_{J_1,J_2,q,p}(w)
  =E_{23}(z)V_{J_1,J_2,q,p}(w)=0.
\eeqa
Note that $p$ and $\beta$ are related to each other as is clear from
(\ref{cartan-gl22}). The conformal dimension of this field is
\beq
\Delta_{J_1,J_2,q,p}=\frac{1}{k}\lt[(J_1-J_2)(J_1+J_2+1)+q(p-2)
  +q^2\lt(\frac{2}{k}-\beta\rt)\rt].\label{dimension}
\eeq

Some remarks are in order. Two arbitrary parameters $q,p$ (or $\beta$) appear
in the vertex operator. This means that to given $J_1,J_2$ which specify
a representation there correspond infinite families of primary
fields. This is yet another interesting characterization
of the non-semisimple $gl(2|2)_k$ current superalgebra. Moreover,
as can be seen from (\ref{dimension}) there exist infinite numbers of
primary fields whose conformal dimensions are negative, which renders 
the CFT non-unitary.

It follows from the results obtained in \cite{Zha04b} that if 
$q\neq\pm (J_1-J_2), \pm(J_1+J_2+1)$, then the corresponding
representations are typical; when $q=\pm (J_1-J_2)$ or $\pm(J_1+J_2+1)$,
atypical representations arise.
The full set (of the two-parameter family) of $gl(2|2)_k$ primary fields 
labelled by $gl(2|2)$ highest weight $(J_1,J_2,q,p)$ are spanned by multiplets
$\Psi_{\cdots}(z)$ of the even subalgebra $gl(2)\oplus gl(2)$, 
and each set contains at most 16 multiplets of $gl(2)\oplus gl(2)$.  
We shall call such a multiplet the level-$x$ multiplet
if the singular parts of its OPEs with 
$H_1(z),H_2(z),H'(z),H_{ex}(z)$ are propotional to
$2(m_1+x), 2(m_2+x), 2q, 2(p-x)$, respectively. These
``levels" should not be confused with the level (specified by $k$) of
the current superalgebra! Then
the level-0 and level-4 multiplets with $gl(2)\oplus gl(2)$ highest
weights $(J_1,J_2,q,p)$ and $(J_1,J_2,q,p-4)$ respectively are 
\beqa
&&\Psi_{J_1,m_1,J_2,m_2,q;p}(z) =(\gm_{12}(z))^{J_1-m_1}(\gm_{34}(z))
  ^{J_2-m_2}V_{J_1,J_2,q,p}(z),\n
&&~~~~m_1=J_1,J_1-1,\cdots,-J_1,~~m_2=J_2,J_2-1,\cdots, -J_2,\n
&&\Psi_{J_1,m_1,J_2,m_2,q;p-4}(z)\n
&&~~=\psi_{23}(z)\psi_{13}(z)\psi_{24}(z) \psi_{14}(z)
  (\gm_{12}(z))^{J_1-m_1-4}(\gm_{34}(z))^{J_2-m_2-4}
   V_{J_1,J_2,q,p}(z),\n
&&~~~~m_1=J_1-4,J_1-5,\cdots,-(J_1+4),~~m_2=J_2-4,J_2-5,\cdots,-(J_2+4).
  \label{level-04b}
\eeqa
Both multiplets have the dimension $(2J_1+1)(2J_2+1)$.

The four level-1 multiplets with
$gl(2)\oplus gl(2)$ highest weights $(J_1-\frac{1}{2},
J_2-\frac{1}{2},q,p-1), (J_1+\frac{1}{2},J_2-\frac{1}{2},q,p-1), 
(J_1+\frac{1}{2},J_2+\frac{1}{2},q,p-1)$ and
$(J_1-\frac{1}{2},J_2+\frac{1}{2},q,p-1)$, respectively, are given by
\beqa
&&\Psi_{J_1-\frac{1}{2},m_1,J_2-\frac{1}{2},m_2,q;p-1}(z)=[\psi_{14}(z)
  +\frac{1}{2}\gm_{12}(z)\psi_{24}(z)-\frac{1}{2}\psi_{13}(z) \gm_{34}(z)\n
&&~~~~ -\frac{1}{3}\gm_{12}(z)\psi_{23}(z)\gm_{34}(z)]
 (\gm_{12}(z))^{J_1-m_1-\frac{3}{2}}(\gm_{34}(z))
  ^{J_2-m_2-\frac{3}{2}}V_{J_1,J_2,q,p}(z),~~~J_1,J_2\geq\frac{1}{2},\n
&&~~~~m_1=J_1-\frac{3}{2},J_1-\frac{5}{2},\cdots,-(J_1+\frac{1}{2}),~~
  m_2=J_2-\frac{3}{2},J_2-\frac{5}{2},\cdots,-(J_2+\frac{1}{2}),\n
&&\Psi_{J_1+\frac{1}{2},m_1,J_2-\frac{1}{2},m_2,q;p-1}(z)\n
&&~~=\lt[\frac{1}{2} (3J_1+m_1+\frac{5}{2})\gm_{12}(z)\psi_{24}(z)
   -\frac{1}{3}(2J_1+m_1+2) \gm_{12}(z)\psi_{23}(z) \gm_{34}(z)\rt.\n
&&~~~~\lt.-(J_1-m_1-\frac{1}{2})[\psi_{14}(z)-\frac{1}{2}\psi_{13}(z)
   \gm_{34}(z)] \rt]\n
&&~~~~\times   (\gm_{12}(z))^{J_1-m_1-\frac{3}{2}}(\gm_{34}(z))
  ^{J_2-m_2-\frac{3}{2}}V_{J_1,J_2,q,p}(z),~~~J_2\geq\frac{1}{2},\n
&&~~~~m_1=J_1-\frac{1}{2},J_1-\frac{3}{2},\cdots,-(J_1+\frac{3}{2}),~~
  m_2=J_2-\frac{3}{2},J_2-\frac{5}{2},\cdots,-(J_2+\frac{1}{2}),\n
&&\Psi_{J_1+\frac{1}{2},m_1,J_2+\frac{1}{2},m_2,q;p-1}(z)\n
&&~~=\lt[-\frac{1}{4}\lt(
  (3J_1+m_1+\frac{5}{2})(3J_2+m_2+\frac{5}{2})+\frac{1}{3}(J_1-m_1-
  \frac{1}{2})(J_2-m_2-\frac{1}{2})\rt)\rt.\n
&&~~~~\times \gm_{12}(z)\psi_{23}(z)\gm_{34}(z)
  +\frac{1}{2}(J_1-m_1-\frac{1}{2})(3J_2+m_2+\frac{5}{2})\psi_{13}(z)
  \gm_{34}(z)\n
&&~~~~ -\frac{1}{2}(3J_1+m_1+\frac{5}{2})(J_2-m_2-\frac{1}{2})\gm_{12}(z)
  \psi_{24}(z)\n
&&~~~~\lt.  +(J_1-m_1-\frac{1}{2})(J_2-m_2-\frac{1}{2})\psi_{14}(z)\rt]
  (\gm_{12}(z))^{J_1-m_1-\frac{3}{2}}(\gm_{34}(z))
  ^{J_2-m_2-\frac{3}{2}}V_{J_1,J_2,q,p}(z),\n
&&~~~~m_1=J_1-\frac{1}{2},J_1-\frac{3}{2},\cdots,-(J_1+\frac{3}{2}),~~
  m_2=J_2-\frac{1}{2},J_2-\frac{3}{2},\cdots,-(J_2+\frac{3}{2}),\n
&&\Psi_{J_1-\frac{1}{2},m_1,J_2+\frac{1}{2},m_2,q;p-1}(z)\n
&&~~=\lt[\frac{1}{2} (3J_2+m_2+\frac{5}{2})\psi_{13}(z)\gm_{34}(z)
  +\frac{1}{3}(2J_2+m_2+2) \gm_{12}(z)\psi_{23}(z)\gm_{34}(z)\rt.\n
&&~~~~\lt.+(J_2-m_2-\frac{1}{2})[\psi_{14}(z)-\frac{1}{2}
   \gm_{12}(z)\psi_{24}(z)]\rt]\n
&&~~~~\times (\gm_{12}(z))^{J_1-m_1-\frac{3}{2}}(\gm_{34}(z)
  ^{J_2-m_2-\frac{3}{2}}V_{J_1,J_2,q,p}(z),~~~J_1\geq\frac{1}{2},\n
&&~~~~m_1=J_1-\frac{3}{2},J_1-\frac{5}{2},\cdots,-(J_1+\frac{1}{2}),~~
  m_2=J_2-\frac{1}{2},J_2-\frac{3}{2},\cdots,-(J_2+\frac{3}{2}).
  \label{level-1b}
\eeqa
The dimensions for these multiplets are $(2J_1)(2J_2),
(2J_1+2)(2J_2), (2J_1+2)(2J_2+2)$ and $(2J_1) (2J_2+2)$, respectively.

The six level-2 multiplets with
$gl(2)\oplus gl(2)$ highest weights $(J_1,J_2-1,q,p-2),
(J_1-1,J_2,q,p-2), (J_1+1,J_2,q,p-2), (J_1,J_2+1,q,p-2),
(J_1,J_2,q,p-2)$ and $(J_1,J_2,q,p-2)$, respectively, are
\beqa
&&\Psi_{J_1,m_1,J_2-1,m_2,q;p-2}(z)=\psi_{24}(z)\psi_{14}(z)
  (\gm_{12}(z))^{J_1-m_1-2}(\gm_{34}(z))^{J_2-m_2-3}V_{J_1,J_2,q,p}(z)\n
&&~~~~+\frac{1}{2}\lt[-\psi_{23}(z)\psi_{14}(z)
  +\frac{1}{6}\psi_{23}(z)\psi_{24}(z)\gm_{12}(z)+\psi_{13}(z)\psi_{24}(z)
  +\frac{1}{2}\psi_{23}(z)\psi_{13}(z)\gm_{34}(z)\rt]\n
&&~~~~\times(\gm_{12}(z))^{J_1-m_1-2}(\gm_{34}(z))^{J_2-m_2-2}
  V_{J_1,J_2,q,p}(z),~~~J_2\geq 1,\n
&&~~~~  m_1=J_1-2,J_1-3,\cdots, -(J_1+2),~~
  m_2=J_2-3,J_2-4,\cdots,-(J_2+1),\n
&&\Psi_{J_1-1,m_1,J_2,m_2,q;p-2}(z)=\psi_{13}(z)\psi_{14}(z)
  (\gm_{12}(z))^{J_1-m_1-3}(\gm_{34}(z))^{J_2-m_2-2}V_{J_1,J_2,q,p}(z)\n
&&~~~~+\frac{1}{2}\lt[\psi_{23}(z)\psi_{14}(z)
  +\frac{1}{6}\psi_{23}(z)\psi_{13}(z)\gm_{34}(z)+\psi_{13}(z)\psi_{24}(z)
  +\frac{1}{2}\psi_{23}(z)\psi_{24}(z)\gm_{12}(z)\rt]\n
&&~~~~\times(\gm_{12}(z))^{J_1-m_1-2}(\gm_{34}(z))^{J_2-m_2-2}
  V_{J_1,J_2,q,p}(z), ~~~J_1\geq 1,\n
&&~~~~m_1=J_1-3,J_1-4,\cdots, -(J_1+1),~~  
  m_2=J_2-2,J_2-3,\cdots,-(J_2+2),\n
&&\Psi_{J_1+1,m_1,J_2,m_2,q;p-2}(z)\n
&&~~=\lt[\frac{1}{2}[J_1-m_1-1+(3J_1+m_1+3)(3J_1+m_1+5)]
  \psi_{23}(z)\psi_{24}(z) (\gm_{12}(z))^2\rt.\n
&&~~~~  +(J_1-m_1-1)(J_1-m_1-2)[\psi_{13}(z)\psi_{14}(z)+
  \frac{1}{12}\gm_{12}(z)\psi_{23}(z)\psi_{13}(z)\gm_{34}(z)]\n
&&~~~~\lt.-\frac{1}{2}(J_1-m_1-1)(3J_1+m_1+4)\gm_{12}(z)[\psi_{13}(z)
  \psi_{24}(z) +\psi_{23}(z)\psi_{14}(z)]\rt]\n
&&~~~~\times (\gm_{12}(z))^{J_1-m_1-3}
  (\gm_{34}(z))^{J_2-m_2-2}V_{J_1,J_2,q,p}(z),\n
&&~~~~m_1=J_1-1,J_1-2,\cdots, -(J_1+3),~~m_2=J_2-2,J_2-3,\cdots,-(J_2+2),\n
&&\Psi_{J_1,m_1,J_2+1,m_2,q;p-2}(z)\n
&&~~=\lt[\frac{1}{4}[J_2-m_2-1+(3J_2+m_2+3)(3J_2+m_2+5)]
  \psi_{23}(z)\psi_{13}(z)(\gm_{34}(z))^2\rt.\n
&&~~~~+\frac{1}{2}(J_2-m_2-1)(3J_2+m_2+4)[\psi_{23}(z)\psi_{14}(z)
  -\psi_{13}(z)\psi_{24}(z)]\gm_{34}(z)\n
&&~~~~\lt.+(J_2-m_2-1)(J_2-m_2-2)[\psi_{24}(z)\psi_{14}(z)
  +\frac{1}{12}\psi_{23}(z)\psi_{24}(z)\gm_{12}(z)\gm_{34}(z)]\rt]\n
&&~~~~\times(\gm_{12}(z))^{J_1-m_1-2}(\gm_{34}(z))^{J_2-m_2-3}
  V_{J_1,J_2,q,p}(z),\n
&&~~~~m_1=J_1-2,J_1-3,\cdots, -(J_1+2),~~m_2=J_2-1,J_2-2,\cdots,-(J_2+3),\n
&&\Psi_{J_1,m_1,J_2,m_2,q;p-2}^{\bf I}(z)\n
&&~~=(J_2-m_2-2)\psi_{24}(z)\psi_{14}(z)
  (\gm_{12}(z))^{J_1-m_1-2}(\gm_{34}(z))^{J_2-m_2-3}V_{J_1,J_2,q,p}(z)\n
&&~~~~+\lt[\frac{1}{2}(J_2+m_2+2)[\psi_{23}(z)\psi_{14}(z)-\psi_{13}(z)
  \psi_{24}(z)]\rt.\n
&&~~~~\lt. +\frac{1}{12}(J_2-m_2-2)\psi_{23}(z)\psi_{24}(z)\gm_{12}(z)
  -\frac{1}{4}(3J_2+m_2+2)\psi_{23}(z)\psi_{13}(z)\gm_{34}(z)\rt]\n
&&~~~~\times(\gm_{12}(z))^{J_1-m_1-2}(\gm_{34}(z))^{J_2-m_2-2}
  V_{J_1,J_2,q,p}(z),\n
&&~~~~m_1=J_1-2,J_1-3,\cdots,-(J_1+2),~~m_2=J_2-1,J_2-2,\cdots,-(J_2+2),\n
&&\Psi_{J_1,m_1,J_2,m_2,q;p-2}^{\bf II}(z)\n
&&~~=(J_1-m_1-2)\psi_{13}(z)\psi_{14}(z)
  (\gm_{12}(z))^{J_1-m_1-3}(\gm_{34}(z))^{J_2-m_2-2}V_{J_1,J_2,q,p}(z)\n
&&~~~~+\lt[-\frac{1}{2}(J_1+m_1+2)[\psi_{13}(z)\psi_{24}(z)
  +\psi_{23}(z)\psi_{14}(z)]\rt.\n
&&~~~~\lt. +\frac{1}{12}(J_1-m_1-2)\psi_{23}(z)\psi_{13}(z)\gm_{34}(z)
  -\frac{1}{4}(3J_1+m_1+2)\psi_{23}(z)\psi_{24}(z)\gm_{12}(z)\rt]\n
&&~~~~\times (\gm_{12}(z))^{J_1-m_1-2}(\gm_{34}(z))^{J_2-m_2-2}
  V_{J_1,J_2,q,p}(z),\n
&&~~~~m_1=J_1-2,J_1-3,\cdots,-(J_1+2),~~m_2=J_2-1,J_2-2,\cdots,-(J_2+2).
  \label{level-2b}
\eeqa
Notice that the last two multiplets, which have been denoted by
$\Psi_{J_1,m_1,J_2,m_2,q; p-2}^{\bf I}(z)$ and 
$\Psi_{J_1,m_1,J_2,m_2,q; p-2}^{\bf II}(z)$,
respectively, have the same highest weight $(J_1,J_2,q,p-2)$. 
It is easy to see from the above expressions
that $\Psi_{J_1,m_1,J_2,m_2,q;p-2}^{\bf I}(z)\equiv 0$ when
$J_2=0$ and $\Psi_{J_1,m_1,J_2,m_2,q;p-2}^{\bf II}(z)\equiv 0$ when $J_1=0$.

The dimensions for the first four multiplets are $(2J_1+1)(2J_2-1),
(2J_1-1)(2J_2+1), (2J_1+3)(2J_2+1)$ and $(2J_1+1)(2J_2+3)$, respectively. 
The dimension for $\Psi_{J_1,m_1,J_2,m_2,q;p-2}^{\bf I}(z)$ is
$(2J_1+1)(2J_2+1)$ if $J_2\neq 0$ and zero if $J_2=0$. Similarly,
the dimension for $\Psi_{J_1,m_1,J_2,m_2,q;p-2}^{\bf II}(z)$ is
$(2J_1+1)(2J_2+1)$ if $J_1\neq 0$ and zero if $J_1=0$.

Finally, the four level-3 multiplets with
$gl(2)\oplus gl(2)$ highest weights $(J_1-\frac{1}{2},
J_2-\frac{1}{2},q,p-3), (J_1+\frac{1}{2},J_2-\frac{1}{2},q,p-3),
(J_1-\frac{1}{2},J_2+\frac{1}{2},q,p-3)$ and
$(J_1+\frac{1}{2},J_2+\frac{1}{2},q,p-3)$, respectively, read
\beqa
&&\Psi_{J_1-\frac{1}{2},m_1,J_2-\frac{1}{2},m_2,q;p-3}(z)
  =\lt[(\psi_{13}(z)+\frac{1}{2}\gm_{12}(z)\psi_{23}(z)
  \psi_{24}(z)\psi_{14}(z)\rt.\n
&&~~~~\lt.+\frac{1}{2}\psi_{23}(z)\psi_{13}(z)[\psi_{14}(z)
  +\frac{1}{3}\gm_{12}(z)\psi_{24}(z)]\gm_{34}(z)\rt]\n
&&~~~~\times(\gm_{12}(z))^{J_1-m_1-\frac{7}{2}}(\gm_{34}(z))^{J_2-m_2-
  \frac{7}{2}}V_{J_1,J_2,q,p}(z),~~~J_1,J_2\geq \frac{1}{2},\n
&&~~~~ m_1=J_1-\frac{7}{2},\cdots,-(J_1+\frac{5}{2}),~~
  m_2=J_2-\frac{7}{2},\cdots,-(J_2+\frac{5}{2}),\n
&&\Psi_{J_1+\frac{1}{2},m_1,J_2-\frac{1}{2},m_2,q;p-3}(z)\n
&&~~ =\lt[-\frac{1}{2}(3J_1+m_1+\frac{9}{2})\psi_{23}(z)
  \psi_{24}(z)\psi_{14}(z)\gm_{12}(z)\rt.\n
&&~~~~+(J_1-m_1-\frac{5}{2})[-\psi_{24}(z)+\frac{1}{2}\psi_{23}(z)
  \gm_{34}(z)]\psi_{13}(z)\psi_{14}(z)\n
&&~~~~\lt.-\frac{1}{6}(5J_1+m_1+\frac{11}{2})\psi_{23}(z)\psi_{13}(z)
  \psi_{24}(z) \gm_{12}(z)\gm_{34}(z)\rt]\n
&&~~~~\times(\gm_{12}(z))^{J_1-m_1-\frac{7}{2}}(\gm_{34}(z))^{J_2-m_2-
  \frac{7}{2}}V_{J_1,J_2,q,p}(z),~~~J_2\geq \frac{1}{2},\n
&&~~~~ m_1=J_1-\frac{5}{2},\cdots,-(J_1+\frac{7}{2}),~~
    m_2=J_2-\frac{7}{2},\cdots,-(J_2+\frac{5}{2}),\n
&&\Psi_{J_1-\frac{1}{2},m_1,J_2+\frac{1}{2},m_2,q;p-3}(z)\n
&&~~ =\lt[-\frac{1}{2}(3J_2+m_2+\frac{9}{2})\psi_{23}(z)\psi_{13}(z)
  \psi_{14}(z) \gm_{34}(z)\rt.\n
&&~~~~+(J_2-m_2-\frac{5}{2})[\psi_{13}(z)+\frac{1}{2}\psi_{23}(z)
   \gm_{12}(z)]\psi_{24}(z)\psi_{14}(z)\n
&&~~~~\lt.-\frac{1}{6}(5J_2+m_2+\frac{11}{2})\psi_{23}(z)\psi_{13}(z)
  \psi_{24}(z)\gm_{12}(z)\gm_{34}(z)\rt]\n
&&~~~~\times(\gm_{12}(z))^{J_1-m_1-\frac{7}{2}}(\gm_{34}(z))^{J_2-m_2-
  \frac{7}{2}}V_{J_1,J_2,q,p}(z),~~~J_1\geq \frac{1}{2},\n
&&~~~~  m_1=J_1-\frac{7}{2},\cdots,-(J_1+\frac{5}{2}),~~
  m_2=J_2-\frac{5}{2},\cdots,-(J_2+\frac{7}{2}),\n
&&\Psi_{J_1+\frac{1}{2},m_1,J_2+\frac{1}{2},m_2,q;p-3}(z)
  =\lt[\frac{1}{4}\lt((3J_1+m_1+\frac{9}{2})(3J_2+m_2+\frac{9}{2})\rt.\rt.\n
&&~~~~\lt.  -\frac{1}{3}(J_1-m_1-\frac{5}{2})(J_2-m_2-\frac{5}{2})\rt)
  \gm_{12}(z)\psi_{23}(z)\psi_{13}(z)\psi_{24}(z)\gm_{34}(z)\n
&&~~~~-\frac{1}{2}(J_1-m_1-\frac{5}{2})(3J_2+m_2+\frac{9}{2})
  \psi_{23}(z)\psi_{13}(z)\psi_{14}(z)\gm_{34}(z)\n
&&~~~~-\frac{1}{2}(3J_1+m_1+\frac{9}{2})(J_2-m_2-\frac{5}{2})
  \gm_{12}(z)\psi_{23}(z)\psi_{24}(z)\psi_{14}(z)\n
&&~~~~\lt.+(J_1-m_1-\frac{5}{2})(J_2-m_2-\frac{5}{2})\psi_{13}(z)
  \psi_{24}(z)\psi_{14}(z)\rt]\n
&&~~~~\times(\gm_{12}(z))^{J_1-m_1-\frac{7}{2}}(\gm_{34}(z))^{J_2-m_2-
  \frac{7}{2}}V_{J_1,J_2,q,p}(z),\n
&&~~~~m_1=J_1-\frac{5}{2},\cdots,-(J_1+\frac{7}{2}),~~
      m_2=J_2-\frac{5}{2},\cdots,-(J_2+\frac{7}{2}).  \label{level-3b}
\eeqa The dimensions for these multiplets are $(2J_1)(2J_2),
(2J_1+2)(2J_2), (2J_1)(2J_2+2)$ and $(2J_1+2)(2J_2+2)$, respectively.

Summarizing, we have obtained 16 independent multiplets,
(\ref{level-04b}) and (\ref{level-1b})-(\ref{level-3b}), of
$gl(2)\oplus gl(2)$, which span the full set (of the two-parameter family) of
primary fields for the
$gl(2|2)_k$ current superalgebra in the standard basis. 
For generic $q$, these primary fields constitute irreducible typical 
representations of $gl(2|2)$ of dimension $16(2J_1+1)(2J_2+1)$. 

The OPEs of the odd currents with the 16 multiplets can be worked out
by means of the results obtained in \cite{Zha04b}, which will not be 
written down here. From these OPEs, it can be shown that
when $q=\pm(J_1-J_2),\pm(J_1+J_2+1)$, the representations become atypical.
So we have four types of one-parameter family of primary fields associated
with the four types of irreducible atypical representations:

\vskip.1in
\noindent \underline{Case 1: $q=J_1-J_2$}.
\vskip.1in
We first consider the case when  $q=J_1-J_2$ and $J_1\neq J_2$. In this
case, all multiplets  drop out except
\beqa
&&\Psi_{J_1,m_1,J_2,m_2,q;p}(z),~~~~
\Psi_{J_1-1/2,m_1,J_2-1/2,m_2,q;p-1}(z),\n 
&&\Psi_{J_1+1/2,m_1,J_2-1/2,m_2,q;p-1}(z),~~~~~
\Psi_{J_1-1/2,m_1,J_2+1/2,m_2,q;p-1}(z),\n
&&\Psi_{J_1,m_1,J_2,m_2,q;p-2}^{\bf sym1}(z)\equiv J_1
\Psi_{J_1,m_1,J_2,m_2,q;p-2}^{\bf I}(z)
+J_2\Psi_{J_1,m_1,J_2,m_2,q;p-2}^{\bf II} (z),\n
&&\Psi_{J_1-1,m_1,J_2,m_2,q;p-2}(z),~~~~
\Psi_{J_1,m_1,J_2-1,m_2,q;p-2}(z),\n
&&\Psi_{J_1-1/2,m_1,J_2-1/2,m_2,q;p-3}(z).
\eeqa
These surviving
multiplets form a one-parameter ($p$ or $\beta$) family of
irreducible atypical representations of $gl(2|2)$ of 
dimension $8[(2J_1+1)J_2+J_1(2J_2+1)]$. When $J_1=J_2\equiv J$ so that $q=0$,
the only surviving multiplets are 
\beqa
&&\Psi_{J,m_1,J,m_2,0;p}(z),~~~~
\Psi_{J+1/2,m_1,J-1/2,m_2,0;p-1}(z),\n
&&\Psi_{J-1/2,m_1,J+1/2,m_2,0;p-1}(z),\n
&&\Psi_{J,m_1,J,m_2,0;p-2}^{\bf sym1'}(z)\equiv 
\Psi_{J,m_1,J,m_2,0;p-2}^{\bf I}(z)
+\Psi_{J,m_1,J,m_2,0;p-2}^{\bf II} (z),
\eeqa
and they give a one-parameter ($p$ or $\beta$) family of irreducible
atypical representations of dimension $4[(2J+1)^2-1/2]$ if $J\neq 0$ and
the trivial one-dimensional representation if $J=0$ (for which the only
multiplet is $\Psi_{0,0,0,0,0;p}(z)$).

\vskip.1in
\noindent\underline{Case 2: $q=-J_1+J_2$}.
\vskip.1in
Let us introduce
\beq
\Psi_{J,m_1,J,m_2,q;p-2}^{\bf sym2}(z)=\lt\{
\begin{array}{ll}
(J_1+1)\Psi_{J_1,m_1,J_2,m_2,q;p-2}^{\bf I}(z) &\\
{}~~+(J_2+1)\Psi_{J_1,m_1,J_2,m_2,q;p-2}^{\bf II}(z) & {\rm for}~J_1\neq 0,~
  J_2\neq 0,\\
\Psi_{0,0,J_2,m_2,q;p-2}^{\bf I}(z) & {\rm for}~J_1=0,\\
\Psi_{J_1,m_1,0,0,q;p-2}^{\bf II}(z) & {\rm for}~J_2=0.
\end{array}
\rt.
\eeq
Then when $q=-J_1+J_2$, the only surviving multiplets are 
\beqa
&&\Psi_{J_1,m_1,J_2,m_2,q;p}(z),~~~~
\Psi_{J_1+1/2,m_1,J_2+1/2,m_2,q;p-1}(z),\n
&&\Psi_{J_1+1/2,m_1,J_2-1/2,m_2,q;p-1}(z),~~~~
\Psi_{J_1-1/2,m_1,J_2+1/2,m_2,q;p-1}(z),\n
&&\Psi_{J_1,m_1,J_2,m_2,q;p-2}^{\bf sym2}(z),~~~~
\Psi_{J_1+1,m_1,J_2,m_2,q;p-2}(z),\n
&&\Psi_{J_1,m_1,J_2+1,m_2,q;p-2}(z),~~~~
\Psi_{J_1+1/2,m_1,J_2+1/2,m_2,q;p-3}(z),
\eeqa
and they consttute a one-parameter ($p$ or $\beta$) family of
irreducible atypical representations of $gl(2|2)$ of
dimension $8[(J_1+1)(2J_2+1)+(2J_1+1)(J_2+1)]$.

\vskip.1in
\noindent\underline{Case 3: $q=J_1+J_2+1$}.
\vskip.1in

Let us introduce
\beq
\Psi_{J,m_1,J,m_2,q;p-2}^{\bf asym3}(z)=\lt\{
\begin{array}{ll}
J_1\Psi_{J_1,m_1,J_2,m_2,q;p-2}^{\bf I}(z) &\\
{}~~-(J_2+1)\Psi_{J_1,m_1,J_2,m_2,q;p-2}^{\bf II}(z) & {\rm for}~J_1\neq 0,~
  J_2\neq 0,\\
0 & {\rm for}~J_1=0,\\
\Psi_{J_1,m_1,0,0,q;p-2}^{\bf II}(z) & {\rm for}~J_2=0.
\end{array}
\rt.
\eeq
Then only the following multiplets
\beqa
&&\Psi_{J_1,m_1,J_2,m_2,q;p}(z),~~~~
\Psi_{J_1-1/2,m_1,J_2-1/2,m_2,q;p-1}(z),\n
&&\Psi_{J_1+1/2,m_1,J_2+1/2,m_2,q;p-1}(z),~~~~
\Psi_{J_1-1/2,m_1,J_2+1/2,m_2,q;p-1}(z),\n
&&\Psi_{J_1,m_1,J_2,m_2,q;p-2}^{\bf asym3}(z),~~~~
\Psi_{J_1-1,m_1,J_2,m_2,q;p-2}(z),\n
&&\Psi_{J_1,m_1,J_2+1,m_2,q;p-2}(z),~~~~
\Psi_{J_1-1/2,m_1,J_2+1/2,m_2,q;p-3}(z),
\eeqa
remain, and they form a one-parameter ($p$ or $\beta$) family of
irreducible atypical representation of $gl(2|2)$
of dimension $8[(2J_1+1)(J_2+1)+J_1(2J_2+1)]$.

\vskip.1in
\noindent\underline{Case 4: $q=-J_1-J_2-1$}.
\vskip.1in
For this case we introduce
\beq
\Psi_{J,m_1,J,m_2,q;p-2}^{\bf asym4}(z)=\lt\{
\begin{array}{ll}
(J_1+1)\Psi_{J_1,m_1,J_2,m_2,q;p-2}^{\bf I}(z) &\\
{}~~-J_2\Psi_{J_1,m_1,J_2,m_2,q;p-2}^{\bf II}(z) & {\rm for}~J_1\neq 0,~
  J_2\neq 0,\\
\Psi_{0,0,J_2,m_2,q;p-2}^{\bf I}(z) & {\rm for}~J_1=0,\\
0 & {\rm for}~J_2=0.
\end{array}
\rt.
\eeq
Then the only surviving multiplets are given by 
\beqa
&&\Psi_{J_1,m_1,J_2,m_2,q;p}(z),~~~~
\Psi_{J_1-1/2,m_1,J_2-1/2,m_2,q;p-1}(z),\n
&&\Psi_{J_1+1/2,m_1,J_2+1/2,m_2,q;p-1}(z),~~~~
\Psi_{J_1+1/2,m_1,J_2-1/2,m_2,q;p-1}(z),\n
&&\Psi_{J_1,m_1,J_2,m_2,q;p-2}^{\bf asym4}(z),~~~~
\Psi_{J_1,m_1,J_2-1,m_2,q;p-2}(z),\n
&&\Psi_{J_1+1,m_1,J_2,m_2,q;p-2}(z),~~~~
\Psi_{J_1+1/2,m_1,J_2-1/2,m_2,q;p-3}(z),
\eeqa
and they give a one-parameter ($p$ or $\beta$) family of
irreducible atypical representations of $gl(2|2)$ of
dimension $8[(J_1+1)(2J_2+1)+(2J_1+1)J_2]$.

\sect{Screening currents in the standard basis}

An important object in the free field approach to CFTs is the screening
current operator. Screening currents are primary fields of conformal
dimension 1, and their OPEs with the currents and the energy-momentum tensor
are regular up to a singular part which is a total derivative. Thus
integrated screening currents (i.e. screening charges) may be inserted into
correlators without altering the conformal or affine Ward identities.
In \cite{Din03b}, three fermionic screening currents of the first kind
corresponding to the three fermionic simple roots of $gl(2|2)_k$ current
superalgebra were obtained. Note that the set of screening currents
depends on the choice of simple root systems of the underlying superalgebra.
In particular, the numbers of bosonic and fermionic screening currents of
the first kind are equal to the numbers of even and odd simple roots,
respectively \cite{Ras01}.

In this section, we construct two bosonic and one fermionic screening 
current operators of the $gl(2|2)$ CFT in the standard basis. The three 
screening currents we find are
\beqa
S_{12}(z)&=&\lt[\beta_{12}(z)+\frac{1}{2}\psi_{23}(z)\psi_{13}^\dg(z)
   +\frac{1}{2}[\psi_{24}(z)+\frac{1}{6}\gamma_{34}(z)\psi_{23}(z)
   ]\psi_{14}^\dg(z)\rt]s_{12}(z),\n
S_{34}(z)&=&-\lt[\beta_{34}(z)-\frac{1}{2}\psi_{23}(z)\psi_{24}^\dg(z)
   -\frac{1}{2}[\psi_{13}(z)-\frac{1}{6}\gamma_{12}(z)\psi_{23}(z)
   ]\psi_{14}^\dg(z)\rt]s_{34}(z),\n
S_{23}(z)&=&-\lt[\psi_{23}^\dg(z)-\frac{1}{2}\gm_{12}(z)\psi_{13}^\dg(z)
   +\frac{1}{2}\gm_{34}(z)[\psi_{24}^\dg(z)-\frac{1}{3}\gm_{12}(z)
   \psi_{14}^\dg(z)]\rt]s_{23}(z),
\eeqa
where 
\beqa
&&s_{12}(z)=\exp\lt[-\frac{1}{\sqrt{k}}\phi_1(z)\rt],~~~~
s_{34}(z)=\exp\lt[\frac{1}{\sqrt{k}}\phi_2(z)\rt],\n
&&s_{23}(z)=\exp\lt[-\frac{1}{2\sqrt{k}}[\phi'(z)-\phi_1(z)+\phi_2(z)]\rt].
\eeqa
The OPEs with the supercurrents and energy-momentum tensor are given by
\beqa
&&E_{i,i+1}(z)S_{j,j+1}(w)=0,\n
&&H_1(z)S_{j,j+1}(w)=H_2(z)S_{j,j+1}(w)=H'(z)S_{j,j+1}(w)=H_{ex}(z)
    S_{j,j+1}(w)=0,\n
&&E_{i+1,i}(z)S_{j,j+1}(w)=\delta_{ij}\p_w\lt(\frac{k}{z-w}s_{j,j+1}(w)\rt),\n
&&T(z)S_{j,j+1}(w)=\p_w\lt(\frac{1}{z-w}S_{j,j+1}(w)\rt),~~~i,j=1,2,3.
\eeqa

\sect{The Knizhnik-Zamolodchikov equations}

Let $\Psi_\L(z)$ be the primary field found in section \ref{primary-fields}
corresponding to the
$gl(2|2)$ representation $\L\equiv (J_1,J_2,q,p)$. Its conformal dimension
$\D_\L$ is given by (\ref{dimension}). Here we only consider the
holomorphic part. The OPEs of such 
a primary field with the energy-momentum tensor $T(z)$ and the supercurrents 
$E_{lm}(z)$ can be worked out and be written as the form:
\beqa
T(z)\Psi_\L(w)&=&\frac{\D_\L}{z-w}\Psi_\L(w)+\frac{1}{z-w}\p_w\Psi_\L(w),\n
E_{lm}(z)\Psi_\L(w)&=&\frac{M_\L^{(lm)}}{z-w}\Psi_\L(w), \label{OPEs}
\eeqa
where $M_\L^{(lm)}$ is the representation of the $gl(2|2)$ generators
$E_{lm}$ for the field $\Psi_\L(z)$, which was obtained in \cite{Zha04b}. 

Introduce the $N$-point correlation function of the primary fields,
\beq
W_N(\{z_i\})=<\Psi_{\L_1}(z_1)\cdots\Psi_{\L_i}(z_i)\cdots \Psi_{\L_N}(z_N)>.
\eeq
Then from the OPEs (\ref{OPEs}) one may derive the Ward identities
corresponding to the global $gl(2|2)$ and conformal symmetries:
\beq
\sum_{j=1}^N\lt(z_j^{n+1}\p_{z_j}+(n+1)\D_{\L_j}z_j^n\rt)W_N(\{z_j\})=0,
     \label{ward1}
\eeq
where $n=-1,0,+1$ and
\beq
\sum_{j=1}^N\,M_{\L_j}^{(lm)}W_N(\{z_j\})=0.\label{ward2}
\eeq
Up to a numerical factor, the two- and three-point  correlation
functions of the $gl(2|2)$ primary fields are
determined by the projective Ward identities (\ref{ward1}). For the 
four-point correlator, the projective Ward identities determine its
form up to a matrix function $F(x)$ of the cross ratio $x=(z_1-z_2)(z_3-z_4)
/ (z_2-z_3)(z_4-z_1)$. 

From the OPEs (\ref{OPEs}) one may derive a set of partial differential
equations that the correlation function $W_N(\{z_i\})$ satisfies. Such
differential equations are known as the Knizhnik-Zamolodchikov (KZ)
equations. It can be shown that the KZ equation for the correlation
function $W_N(\{z_i\})$ has the form,
\beq
\lt[k\p_{z_i}-\sum_{j\neq i}\frac{\frac{(-1)^{[m]}}{2}\lt(M_{\L_i}^{(lm)}
   M_{\L_j}^{(ml)}+M_{\L_j}^{(lm)}M_{\L_i}^{(ml)}\rt)+\frac{q^2}{k}}{z_i-z_j}
   \rt]W_N(\{z_i\})=0,\label{KZ}
\eeq
where it should be understood that both indices $l$ and $m$ are summed 
from 1 to 4.
For the four-point correlator, the KZ equation reduces to a matrix
differential equation for $F(x)$. Further computation depends on the
choice of the representations of the primary fields labelled by $\L_i$, 
which is beyond the scope of the present paper. A detailed discussion of
this and other related problems concerning the $gl(2|2)$ CFT will be
given in a separate paper.

\sect{Conclusions and discussions}

In this paper we have continued our investigation of current
superalgebras and their corresponding non-unitary CFTs.
We have constructed the Sugawara energy-momentum tensor, an infinite
family of primary fields and three screening current operators of 
the non-unitary $gl(2|2)$ CFT in the standard basis for general level $k$. 

As mentioned in the introduction, one of the motivations for this study 
is the application of the current superalgebra and the corresponding
non-unitary CFT to disordered systems such as the integer quantum Hall
transition. The results obtained in this paper provide some basic
ingredients useful in the supersymmetric treatment \cite{Efe83}
of certain Gaussian type disordered systems. Such applications are 
under investigation, and results will be presented elsewhere.

\vskip.1in

\no {\bf Acknowledgments:}
This work was financially supported by the Australian Research Council.
One of the authors (Liu) has been supported by IPRS and UQGSS scholarships
of the University of Queensland.

\bebb{99}

\bbit{Roz92} 
L. Rozanski and H. Saleur, \npb {376} {1992} {461}.

\bbit{Isi94}
J. M. Isidro and A. V. Ramallo, \npb {414}{1994}{715}. 

\bbit{Flohr}
M. Flohr, Int. J. Mod. Phys. {\bf A28}, (2003) 4497.

\bbit{Ber95}
D. Bernard, eprint hep-th/9509137.

\bbit{Mud96}
C. Mudry, C. Chamon and X.-G. Wen, \npb {466} {1996} 383.

\bbit{Maa97}
Z. Maassarani and D. Serban, \npb {489} {1997} {603}.

%\bbit{Zir99}
%M.R. Zirnbauer, preprint hep-th/9905054.

\bbit{Bas00}
Z.S. Bassi and A. LeClair, \npb {578} {2000} {577}.

\bbit{Gur00}
S. Guruswamy, A. LeClair and A.W.W. Ludwig, \npb {583} {2000} {475}.

%\bbit{Bha01}
%M. J. Bhaseen,  J.-S. Caux ,  I. I. Kogan  and  A. M. Tsvelik, 
%\npb {618}{2001}{465}.

\bbit{Ber99} M. Bershadsky, S. Zhukov and A. Vaintrob, 
\npb {559} {1999} {205}.

\bbit{Fra96}
L. Frappat, P. Sorba and A. Sciarrino, ``Dictionary on Lie
superalgebras", eprint hep-th/9607161.

\bbit{Ito}
K. Ito, \plb {252}{1990}{69}.

\bbit{Bow96}
P. Bowcock, R-L.K. Koktava and A. Taormina, \plb {388} {1996} {303}.

\bbit{Ras01}
J. Rasmussen, \npb {510} {1998} {688}; \npb {593} {2001} {634}.

\bbit{Din03a}
X.M. Ding, M. D. Gould, C. J. Mewton and Y. Z. Zhang, 
J. Phys. {\bf A36}, (2003) 7649.

\bbit{Zha03a}
Y.Z. Zhang, Phys. Lett. {\bf A327}, (2004) 442.

\bbit{Din03b}
X.M. Ding, M. D. Gould and Y. Z. Zhang, Phys. Lett. {\bf A318}, (2003) 354.

\bbit{Zha04b}
Y.Z. Zhang and M.D. Gould, eprint math.QA/0405043, J. Math. Phys., in press.

\bbit{Efe83}
K. Efetov, Adv. Phys. {\bf 32}, (1983) 53. 
\eebb

\end{document}